\documentclass[reprint,superscriptaddress,amsmath,amssymb,aps,pre]{revtex4-2}
\usepackage{xr}
\usepackage[normalem]{ulem}
\usepackage{hyperref}
\usepackage{gibbon}
\usepackage{xcolor}
\usepackage[utf8]{inputenc}
\usepackage{cleveref}
\begin{document}

\preprint{APS/123-QED}
\title{Insights from a pseudospectral study of a potentially singular solution
of the three-dimensional axisymmetric incompressible Euler equation}
\author{Sai Swetha Venkata Kolluru}
\email{saik@iisc.ac.in}
\affiliation{Centre for Condensed Matter Theory, Department of Physics, Indian Institute of Science, Bangalore, 560012, India. }
\author{Puneet Sharma}
\email{puneet.sharma@ds.mpg.de}
\affiliation{Dynamics of Complex Fluids (DCF), Max Planck Institute for Dynamics and Self-Organization, Am Fassberg 17, 37077 G\"{o}ttingen, Germany}
\author{Rahul Pandit}
\email{rahul@iisc.ac.in}
\affiliation{Centre for Condensed Matter Theory, Department of Physics, Indian Institute of Science, Bangalore, 560012, India}
\date{\today}

\begin{abstract}
We develop a Fourier-Chebyshev pseudospectral direct numerical simulation (DNS)
to examine a potentially singular solution of the radially bounded,
three-dimensional (3D), axisymmetric Euler equations [G. Luo and T.Y.
Hou, Proc. Natl. Acad. Sci. USA, 111.36 (2014)]. We demonstrate that:
(a) the time of singularity is preceded, in any spectrally truncated
DNS, by the formation of oscillatory structures called tygers, first
investigated in the one-dimensional (1D) Burgers and two-dimensional (2D) Euler equations; (b) the
analyticity-strip method can be generalized to obtain an estimate for
the (potential) singularity time.
\end{abstract}

\pacs{}                            
\maketitle

\section{Introduction}
Two hundred and sixty five years ago, Euler introduced the equations for an
inviscid, incompressible, three-dimensional (3D) fluid in \textit{Principes
g\'en\'eraux du mouvement des fluides}~\cite{euler1755,euler2008,aussois}.  The
incompressible Euler partial differential equation (PDE) and its descendant,
the incompressible Navier-Stokes PDE~\cite{navier1822,stokes1880}, govern,
respectively, ideal and viscous fluid flows at low Mach numbers. They are, therefore, among the most prominent equations in physics; and their solutions are of importance in a variety of physical settings.  Furthermore, these
equations  pose challenges for mathematicians: It is well known that the solutions of the two-dimensional(2D) Euler equation, with analytic initial data, do not exhibit a finite-time singularity \cite{2006pauls}; however, it is still not known if any solutions of the 3D Euler equations develop a singularity in a finite time, if we start with
analytic initial data (for non-analytic initial data, see Ref.~\cite{elgindi2019Euler3D}).
The answer to this grand-challenge, finite-time-singularity problem also has important implications for turbulence in fluids, even if we use the 3D Euler PDE, as conjectured by Onsager~\cite{onsager,constantin1994onsager}; for a detailed discussion of these issues, see, e.g., Refs.~\cite{eyink,eyink2006onsager}, and, for recent advances, Ref.~\cite{buckmaster2021convex}. 

The possible relation between finite-time-singularities in the 3D Euler PDE and finite-dissipation weak solutions of the 3D Euler equations, and their potential relevance to solutions of the 3D Navier-Stokes equation in the limit of vanishing viscosity are discussed in  Refs.~\cite{eyink,eyink2006onsager,de2010admissibility,de2012h,de2013dissipative,de2014dissipative}. In this paper, we do not address the regularity problem for the 3D Navier-Stokes PDE, which is one of the Clay Mathematics problems; for a discussion of this problem we refer the reader to Ref.~\cite{clay}.
Here, we investigate a potentially singular solution, first studied by Luo and Hou~\cite{houluo}, of a 3D axisymmetric Euler flow.

Explorations of finite-time-singularity problems (for the Euler case see, e.g., Refs.~\cite{aussois,gibbonrev}) often use direct numerical simulations (DNSs), which have not yielded unambiguous results for or
against a finite-time singularity in the 3D Euler PDE.  Luo and Hou~\cite{houluo} have explored a \textit{potentially singular solution of the radially bounded, 3D, axisymmetric Euler equations}
via a hybrid Galerkin and finite-difference method. Given the importance of this problem, it behooves us to study this potentially singular solution by a completely different numerical scheme and another singularity-detection criterion, in addition to the one based on the well-known Beale-Kato-Majda theorem~\cite{houluo,beale1984remarks,bkmas}. In particular, we use the singularity-detection criterion based on the movement of singularities in the complex space that was first discussed in the work of Sulem, \textit{et al.}~\cite{sulem,kida1986study,ootb,bkmas,cickptg}. This method, referred to as the analyticity-strip method, calls for a pseudo-spectral simulation of the governing PDEs.

Therefore, we have developed a pseudospectral, Fourier-Chebyshev scheme to study this problem; in
any numerical implementation, we can only use a \textit{finite number} of Fourier-Chebyshev modes, i.e., we have a spectrally truncated system. 

Our method leads to new insights that include the formation of localized,
oscillatory structures, called \textit{tygers}, at points of positive strain in the velocity fields. Tygers were first introduced in the context of the one-dimensional (1D) Burgers and two-dimensional (2D) Euler equations~\cite{tyger1,tyger2,di2018dynamics,banerjee2014transition,pramana}, \textit{en route} to \textit{thermalization}, in spectrally truncated pseudospectral DNSs; note that the appearance of tygers does not necessarily imply the formation of a finite-time singularity, which occurs in the inviscid 1D Burgers equation but not for the 2D Euler PDE.
Lee~\cite{lee1952some} and Hopf~\cite{hopf1952statistical} had proposed~\cite{kraichnan1955statistical,cartes2021galerkin} that such spectrally truncated systems, with a finite number of modes, \textit{must thermalize, at sufficiently long times}, because the total energy is conserved; the thermalized state displays equipartition of the energy between all wavenumber ($k$) modes. Such thermalization has been observed in various spectrally truncated hydrodynamical equations including the 3D Euler~\cite{cichowlas2005effective} and the 3D and 2D Gross-Pitaevskii~\cite{krstulovic2011energy,shukla2013turbulence} equations. 
The high-$k$ modes thermalize faster than the low-$k$ ones in, e.g., the spectrally truncated 3D Euler equation; these high-$k$ thermalized modes act effectively as a dissipation range for the low-$k$ modes and, over intermediate time scales, before complete thermalization occurs, the fluid energy spectrum shows a power law $\sim k^p$ form with the exponent $p \simeq -5/3$ as in the Kolmogorov 1941 phenomenology for inertial-range scaling in 3D Navier-Stokes (NS) turbulence~\cite{cichowlas2005effective}. We note, in passing, that high-order hyperviscosity in the 3D NS equation can emulate these effects of Galerkin truncation in the 3D Euler PDE as discussed in Ref.~\cite{frisch2008hyperviscosity}. A discussion of hyperviscosity is out of place here because we are concentrating on the 3D axisymmetric Euler PDE; a full discussion of Galerkin truncation via very-high-order hyperviscosity would require a separate study. 

We concentrate on the Galerkin-truncated axisymmetric 3D Euler PDE. We find that, before the appearance of tygers, our method yields spectral convergence to the 3D Euler PDE we consider, and the truncated solution is the true solution; soon after the birth of tygers, our spectrally truncated system moves towards thermalization and it does not provide a good representation of this PDE. Nevertheless, we show how to generalize the analyticity-strip method to uncover signatures of the potential singularity discussed above. 

The remainder of this paper is organised as follows: In Sec.~\ref{sec:Model} we define the model we study. Section~\ref{sec:NM} contains the numerical methods we use.
In Sec.~\ref{sec:Results} we present the results of our study. Section~\ref{sec:Conclusions} contains a discussion of our results in the light of earlier studies. Some details of our calculations are given in the Appendices~\ref{app:rescomp}-~\ref{app:stat}.

\section{Model}
\label{sec:Model}

\begin{figure}[t]
\centering
\includegraphics[width=0.9\linewidth]{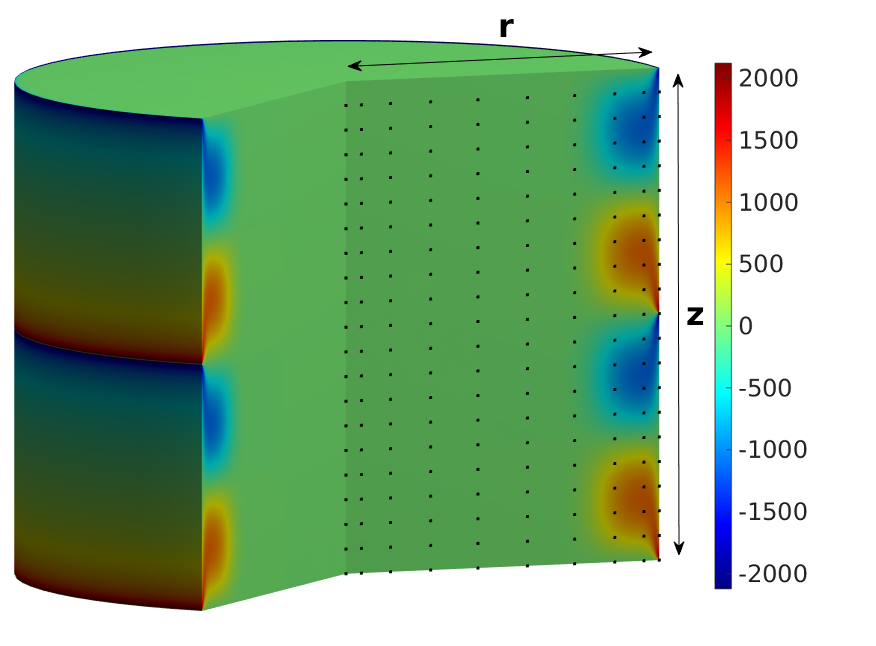}
\caption{(Color online) A section of our cylindrical simulation domain with the heat-map of $\omega^1$ at a representative time $t=0.003094$ for a resolution of $N_r= 512$ and $N_z= 1024$. Chebyshev collocation points are shown schematically in the $r-z$ plane for a constant value of $\theta$; these are spaced more closely near $r=0$ and $r=1$ than in the middle of the domain.}
\label{fig:cyl}
\end{figure}
The 3D Euler PDE, for an incompressible, inviscid fluid is 
\begin{eqnarray}
\bom_t + \bu \cdot \nabla \bom &=& \bom \cdot \nabla \bu ; \nonumber \\
\bom = \nabla \times \bu ; \; \bu &=& \nabla \times \bpsi ;
\label{eq:3DEuler}
\end{eqnarray}
here, $\bom$ is the vorticity, $\bu$ the velocity field, and
$\bpsi$ the vector-valued stream function that is related to
the vorticity by the Poisson equation $\bom = - \nabla^2 \bpsi$; and $\bom_t \equiv \partial \bom /\partial t$.
For axisymmetric flows, we use $
\bu(r,z) = u^r (r,z) \ \mathbi{\hat{e}_r} + u^{\theta}(r,z)  \
\mathbi{\hat{e}_{\theta}} + u^z(r,z)  \ \mathbi{\hat{e}_z} $, where $ \
\mathbi{\hat{e}_r} , \ \mathbi{\hat{e}_{\theta}}$, and $ \ \mathbi{\hat{e}_z}$
are unit vectors in the cylindrical coordinate system.
Then, Eq.\eqref{eq:3DEuler} can be reduced to a system of equations for 
\begin{equation}
u^1 = u^{\theta}/r, \qquad  \omega^1 = \omega^{\theta}/r, \qquad \psi^1=\psi^{\theta}/r,
\label{eq:u1etc}
\end{equation}
where $u^{\theta}$, $\omega^{\theta}$, and $\psi^{\theta}$ are angular components:
\begin{subequations}
\label{eq:set}
	\begin{align}
	u^1_t + u^ru^1_r + u^zu^1_z &= 2u^1\psi_z^1 ,\label{eq:main1}\\
	\omega_t^1 + u^{r}\omega_r^1 + u^z\omega_z^1 &= ((u^1)^2)_z ,\label{eq:main2}\\
	-\Big( \partial_r^2 + \frac{3}{r}\partial_r + \partial_z^2 \Big) \psi^{1} &= \omega^{1} ,
    \label{eq:main3}
	\end{align}
	with $u^{r} = - r\psi^{1}_{z}$ and $u^{z} = 2\psi^{1}+r\psi^{1}_{r}$; and the subscripts $r, \, t,$ and $z$ on the functions indicate $\partial_r$, $\partial_t$, and $\partial_z$, respectively.
\label{eq:AxisymmetricEuler}
\end{subequations}

The variables $u^1, \omega^1$, and $\psi^1$ are well defined, so long as the
solutions to Eq.\eqref{eq:AxisymmetricEuler} are smooth 
{($C^{\infty}(\mathbf{R} \times \bar{\mathbf{R}}^{+})$} with $\mathbf{R}$, the set of real numbers and $\bar{\mathbf{R}}^{+}$, the set of affinely extended positive real numbers); $u^{\theta}, \omega^{\theta}$, and
$\psi^{\theta}$ must all vanish at $r=0$ for these solutions to remain
smooth~\cite{liuwang}. We solve Eq.\eqref{eq:AxisymmetricEuler} in the domain $
D(1,L) = \{ (r,z) : 0 \leq r \leq 1 , 0 \leq z \leq L \};$ we use $L$-periodic
boundary conditions in $z$, the no-flow condition at $r=1$ \eqref{eq:noflow},
and the pole condition at $r=0$ \eqref{eq:polecond}: 
\begin{eqnarray}
		\psi^{1}(r=1,z,t) &=& 0; \label{eq:noflow} \\
	u^{1}_r(r=0,z,t)= \omega^{1}_r(r=0,z,t)
	&=&\psi^{1}_r(r=0,z,t) \nonumber \\
	&=& 0;
	\label{eq:polecond}
\end{eqnarray}
and the initial data \cite{houluo}:
\begin{subequations}
    \begin{align}
    u^{1}(r,z,t=0)& = 100e^{-30(1-r^{2})^{4}}\sin\Big({\frac{2\pi z}{L}\Big)}; \\
    \omega^{1}(r,z,t=0)& = \psi^{1}(r,z) = 0.
    \label{eq:initial3D}
    \end{align}
\end{subequations}

To compare our results with those of Luo and Hou~\cite{houluo}, it is \textit{imperative} that we use their initial condition. (See Appendix \ref{app:stat} for other types of initial conditions.)

\begin{figure}[h!]
\centering
\includegraphics[width=\linewidth]{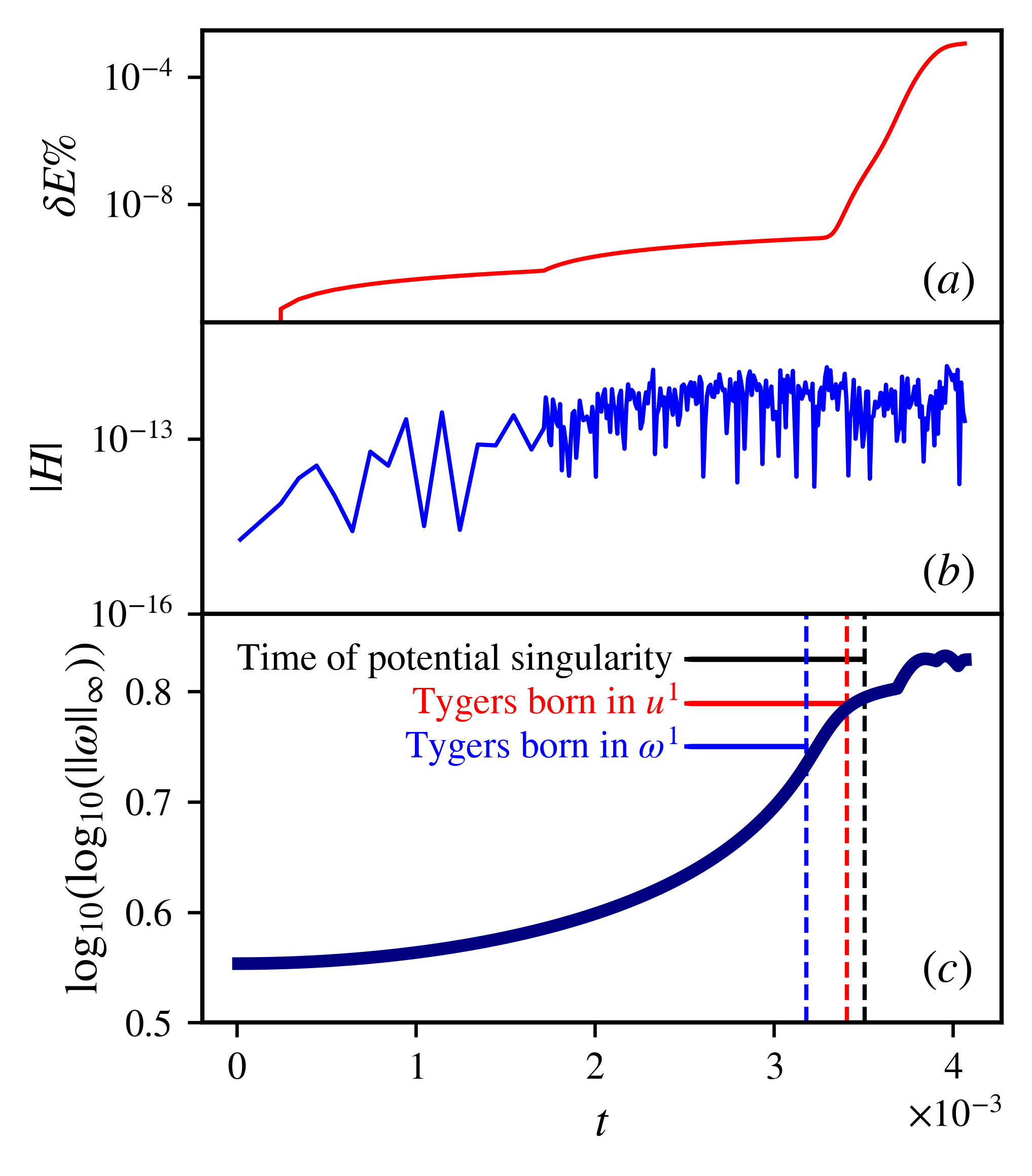}
\caption{(Color online){ Plots versus $t$ of $(a)$
log (base 10) of the percentage change, in our DNS, of the energy $ (\delta
E \%)$ (red full line), $(b)$ log (base 10) of the absolute value $|H|$
of the helicity (Eq.~\ref{eq:eh_exp}), and $(c)$ $\log_{10}(\log_{10}(||\omega||_{\infty}))$ (dark blue full line) for $N_z = 4096$ and $N_r = 512$. Here,
$||\omega||_{\infty}$, the $L_\infty$ norm of the vorticity, is well approximated by
the maximum value of $|\omega|$ on our grid. The red (blue) dashed line
indicates the time of the birth of a tyger (see text) in $u^1$ ($\omega^1$);
the black dashed line denotes the estimate for the time of the (potential)
singularity, from Ref.~\cite{houluo}. In Fig. \ref{fig:2res} of Appendix \ref{app:rescomp}, we give similar plots for other values of $N_z$ and $N_r$; the
higher the values of  $N_z$ and $N_r$ (especially $N_z$), the better our scheme
captures the rapid growth of $\log_{10}(\log_{10}(||\omega||_{\infty}))$.}}
\label{fig:main_1}
\end{figure}
\section{Numerical methods} 
\label{sec:NM}

\subsection{Fourier Chebyshev spectral methods}

\begin{figure*}
\centering
    \includegraphics[width=\linewidth]{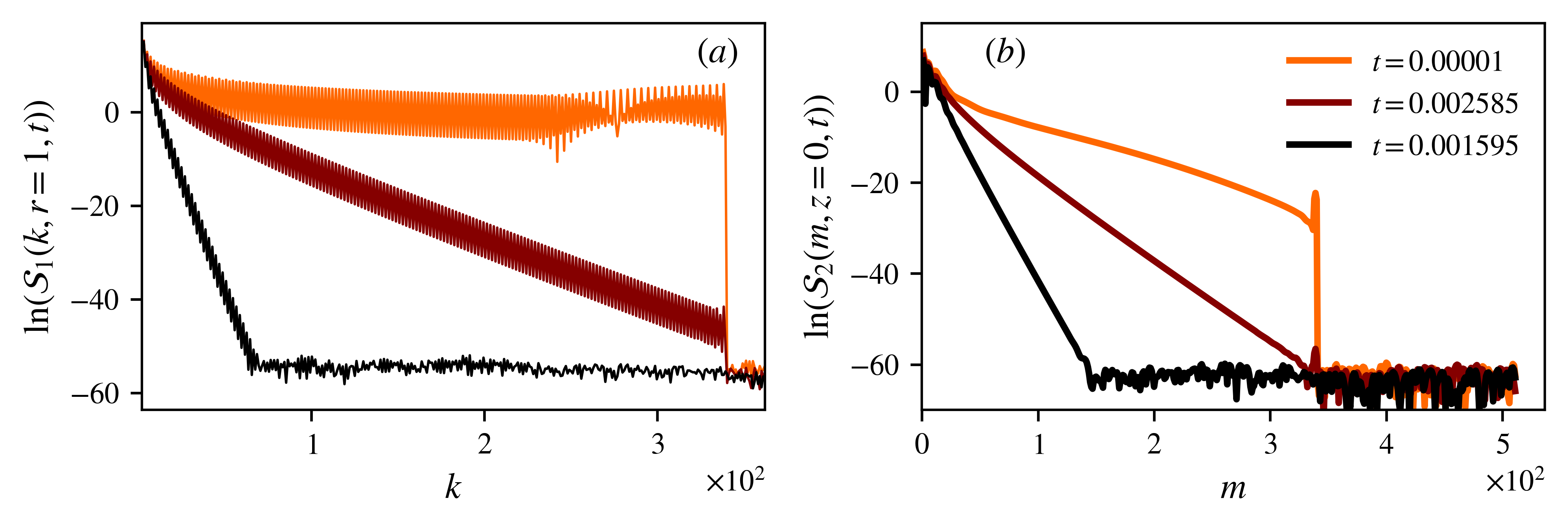}
\caption{(Color online) $(a)$ Plots versus $k$ of $\ln(\mathcal{S}_1(r=1,k,t))$, at different times $t$ (the full temporal evolution is given in the video S1 in the Supplemental Material \cite{supp}); here, the modes with $k > k_G$, the dealiasing-cutoff wavenumber, have zero energy. $N_r = 512$, $N_z=1024$, and the dealiasing cutoff is $k_G=341$. $(b)$ Plots versus $m$ of $\ln(\mathcal{S}_2(m,z=0,t))$ at different times $t$; there is an exponentially decaying tail in the spectrum $\mathcal{S}_2(m,z=0,t)$, at large $m$, whose decay rate decreases with $t$. (The full temporal evolution is given in the video S2 in the Supplemental Material \cite{supp}.) }
\label{fig:main_2}
\end{figure*}

We use the Fourier-Chebyshev representation, in which a function $f(r,z)$ is approximated by
\begin{equation}
f(r,z) = 
\sum_{k}  \sum_{m}\hat{f}(k,m)  e^{ikz} \ T_m(2r-1),
\end{equation}
where $T_m$ is the Chebyshev polynomial (of the first kind) of order $m$.
{
In the schematic diagram in Fig.~\ref{fig:cyl}, we display the collocation
points in our Fourier-Chebyshev DNS; these points are distributed uniformly in the periodic (axial) direction $z$; in the radial direction $r$, these points coincide with the roots of the highest-order Chebyshev polynomial in our basis. 
We use a finer resolution in the $z$ direction than in the $r$ direction, because, for a given number of collocation points, the Chebyshev nodes are spaced more closely near the boundary at $r=1$ than the Fourier nodes. This prevents excessive elongation of the cells in our simulation grid, in physical space near this boundary. If these cells are very elongated and narrow in the radial direction, it becomes difficult to satisfy the Courant-Friedrichs-Lewy (CFL) condition at every time-integration step. 
We use a CFL number $C =0.2$ and adjust the time step $dt$, to ensure that the CFL condition is satisfied. For the temporal evolution of Eqs.\eqref{eq:main1},\eqref{eq:main3}, we use the explicit fourth-order Runge-Kutta scheme in physical space; we evaluate the derivatives in Fourier-Chebyshev space and, subsequently, compute the nonlinear terms in physical space. 
We solve the Poisson equation Eq.\eqref{eq:main3} in the domain $D(1,L) = \{ (r,z) : 0 \leq r \leq 1 , 0 \leq z \leq L \}$ with the boundary conditions Eq.\eqref{eq:noflow}, \eqref{eq:polecond}. 
We use the $2/3$ truncation method for dealiasing both Fourier and Chebyshev modes. 
Reference~\cite{houluo} utilizes the symmetry properties of this initial condition  to study the Euler PDEs in the domain $\mathcal{D}(1,L/4)$; in our Fourier-Chebyshev method we use the full length $L$ of the domain. }

\subsection{Conserved quantities and Spectra}

The total energy and helicity are, respectively,
\begin{subequations}
\begin{align}
E &= \frac{1}{2} \int^1_0 \int^L_0 (|u^r|^2 +|u^z|^2 
+ |u^{\theta}|^2) \ r dr dz ; \\
H &=\int^1_0 \int^L_0 \bu \cdot \bom \ r dr dz .
\label{eq:eh_exp}
\end{align}
\end{subequations}
We calculate these by using the Fourier-Chebyshev coefficients of $\bu$ and $\bom$ 
{(see Figs. \ref{fig:main_1}(a) and (b))}.

Fourier and Chebyshev transforms, over $z$ and $r$, respectively, yield the fixed-$r$ and fixed-$z$ spectra
\begin{subequations}
\begin{align}
\mathcal{S}_1(r,k,t) &:= \frac{g(k)}{2 \ N_z} \Big( |\hat{u}^{\theta}(r,k,t)|^2 +  \nonumber \\ 
	                & |\hat{u}^{r}(r,k,t)|^2 + |\hat{u}^{z}(r,k,t)|^2 \Big),  \label{eq:S1} \\	                
\mathcal{S}_2(m,z,t) &:= \frac{N_r}{2 \ g(m)} \Big( |\hat{u}^{\theta}(m,z,t)|^2 + \nonumber \\
                    &  |\hat{u}^{r}(m,z,t)|^2 +  |\hat{u}^{z}(m,z,t)|^2 \Big), \label{eq:S2}
\end{align}
\end{subequations}
where $g(i=0) = 1$ and $g(i>0)=2$ ($i$ is $k$ or $m$). 
We give the spatiotemporal evolution of $\mathcal{S}_1(r,k,t)$ and $\mathcal{S}_2(m,z,t)$ in videos S1 and S2, respectively, in the Supplemental Material \cite{supp}. 
  Similarly, simultaneous Fourier-Chebyshev transforms give us the following spectra
\begin{subequations}
\begin{align}
\mathcal{S}_3(m,k,t) := \Big(& |\hat{u}^{\theta}(m,k,t)|^2 +  \nonumber \\ 
	                & |\hat{u}^{r}(m,k,t)|^2 + |\hat{u}^{z}(m,k,t)|^2 \Big),  \label{eq:S3} \\	            
\mathcal{S}_4(m,k,t) := \Big(&  |\hat{u}^{\theta}(m,k,t) \ \hat{\omega}^{\theta}(m,k,t)| +  \nonumber \\ 
	                & |\hat{u}^{r}(m,k,t) \ \hat{\omega}^{r}(m,k,t)| + \nonumber \\ 
	                & |\hat{u}^{z}(m,k,t) \ \hat{\omega}^{z}(m,k,t)| \Big). \label{eq:S4}
\end{align}
\end{subequations}

\subsection{Methods to track singularity}
\subsubsection{Beale-Kato-Majda criterion: Growth of $||\omega||_\infty$}
The detection of a singularity based on the BKM theorem~\cite{beale1984remarks,bkmas} 
uses a plot of $\log_{10}(\log_{10}(||\omega||_{\infty}))$ versus $t$. 
We show such a plot (blue full line) in Fig.~\ref{fig:main_1}$(c)$, from our DNS; the red (blue) dashed line indicates the time of the birth of a tyger (see below)
 in $u^1$ ($\omega^1$); the black dashed line denotes the estimate for the time of
 the (potential) singularity, from Ref.~\cite{houluo}.

\begin{figure*}[t]
\centering
\includegraphics[width=\linewidth]{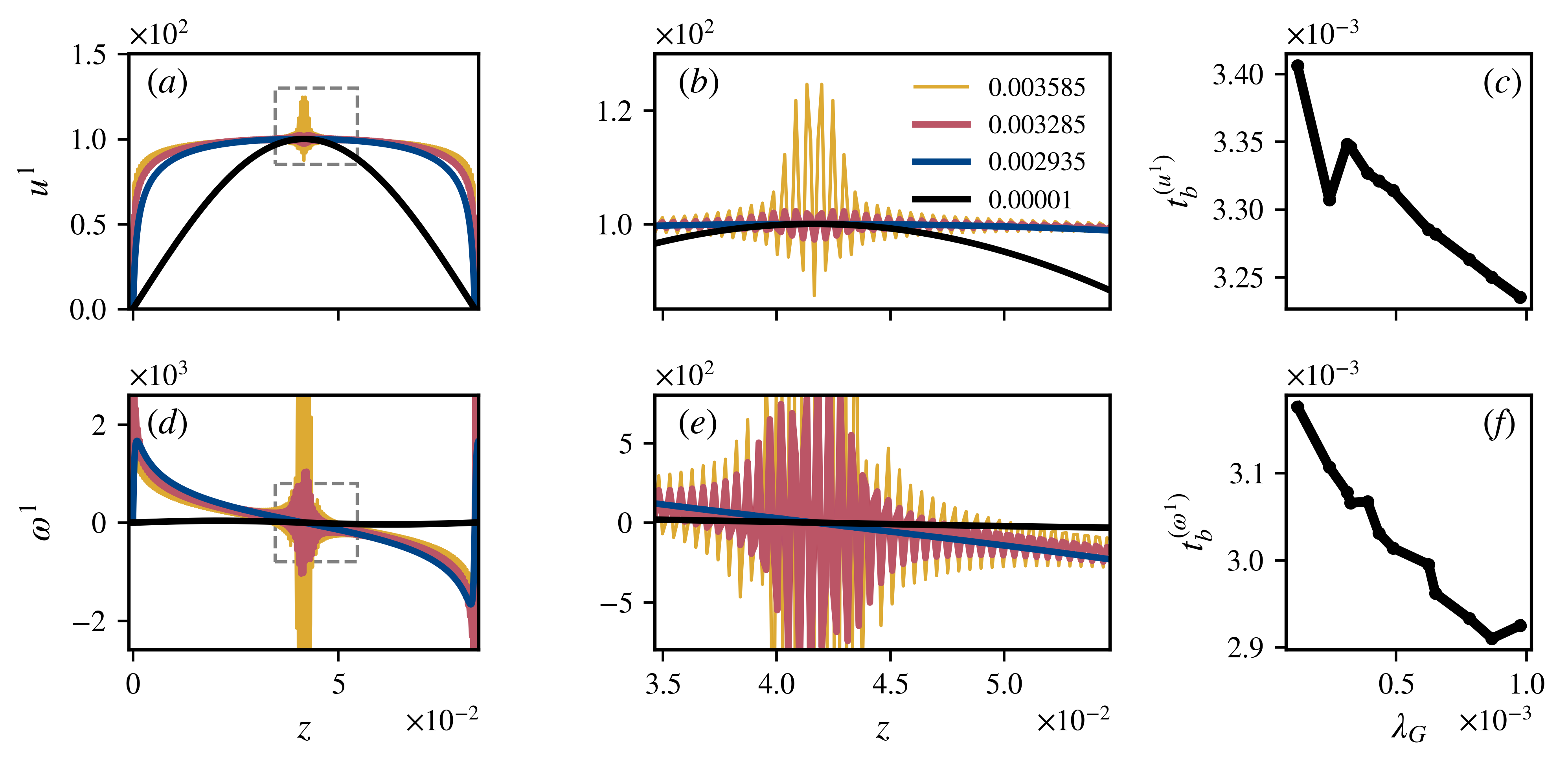}
\caption{(Color online) Plots versus $z$ of $(a),(b)$ $u^{1}(r=1)$ and $(d),(e)$ $\omega^{1}(r=1)$ at various times $t$ listed in panel $(b)$ for $N_r=512$ and $N_z=1024$; as we go from
columns one to two, we zoom in to the region with localized oscillatory structures called \textit{tygers}~\cite{tyger1,tyger2}. Plots of the tyger-birth time $t_{b}$ versus $\lambda_G = 2 \pi/k_G$ in $(c)$ $u^1$ and $(f)$ $\omega^1$, respectively, where $k_G$ is the dealiasing cutoff wavenumber. To determine the tyger-birth times $t_b^{(\bu^1)}$ and $t_b^{(\bom^1)}$), we examine plots of $\bu_1$ and $\bom_1$ as a function of $t$.} 
\label{fig:main_3}
\end{figure*}
\subsubsection{Analyticity Strip method}
For a DNS in a domain with periodic boundary conditions in all spatial directions, the analyticity-strip method~\cite{sulem,brachet1983small,kida1986study,brachet1992numerical,ootb,bkmas,cickptg}
proposes that the solution of the PDE can be continued analytically to complex
space variables $ {\bf z} = {\bf x} + i{\bf y}$, inside the \textit{analyticity
strip} $\mid {\bf y} \mid < \delta(t)$, where $t$ is real and
$\delta(t)$, the width of this strip, follows from the spatial Fourier
transform of the solution, which decays, at large wavenumbers $k$, as $\exp(-k
\delta (t))$ (this has an algebraic prefactor). We obtain $\delta(t)$ and 
estimate if $\delta(t) \to 0$ at a finite time $t^*$; at this time the solution shows a
finite-time singularity because singularities, in the complex plane for $t <
t^*$, hit the real axis. Our determination of $\delta(t)$ is accurate up until
times at which $\delta(t)$ remains larger than a few mesh widths.  For such
times, we have spectral convergence of the Fourier expansion.

We now extend the analyticity-strip method: (a) We first work with a fixed value of $r$; we evaluate the Fourier transform(in the $z$ direction) of the components of the velocity; the wavenumber dependence of this transform yields the width of this analyticity strip.  (b) Next, we work with a fixed value of $z$; we evaluate the Chebyshev transform(in the $r$ direction) of the components of the velocity; we then examine the dependence of the Chebyshev-expansion coefficients~\cite{gargano2009singularity,takeshi,takeshi1,takeshi2,takeshi3,takeshi4} on the order $m$; if these coefficients decrease as $\exp \ (-m \alpha)$, for large $m$, then the velocity field is analytic in the Bernstein ellipse $\mathcal{E}_{\rho_{*}} =\{z\in \mathbb{C} \mid z= (\rho_{*}e^{\imath \theta} - \rho_{*}^{-1}e^{-\imath \theta})/2, 0 \leq \theta \leq 2 \pi \}$, with
\begin{equation}
\rho_* = e^{\alpha};\;    \rm{and} \;\;
\delta_r=( \rho_{*} - \rho_{*}^{-1})/2,
\label{eq:alpha_dr}
\end{equation}
the width of this analyticity strip.

Before the birth of tygers, we have spectral convergence of our Fourier-Chebyshev expansions. This allows us to employ the analyticity-strip method. We concentrate on $\mathcal{S}_1(r,k,t)$ and $\mathcal{S}_2(m,z,t)$.

In Fig.~\ref{fig:main_2}$(a)$ we plot $\ln (\mathcal{S}_1(r=1,k,t))$(see Eq. \ref{eq:S1}) versus $k$,  at different times
$t$ (see the video S1 in the Supplemental Material \cite{supp}); here, the
modes with $k > k_G$, the dealiasing-cutoff wavenumber, have zero energy.       

The symmetries of our initial condition lead to even-odd $k$ oscillations in, e.g., $\mathcal{S}_1(r=1,k,t)$ (black, brown, and orange curves in Fig.~\ref{fig:main_2}$(a)$. At small and intermediate values of $t$, these oscillations have exponentially decaying envelopes at large $k$. The envelope for odd $k$ lies above its even-$k$ counterpart and the separation between these envelopes increases with $t$. The natural logarithmic decrements of these envelopes, $\delta_{\text{odd}}(t)$ and $\delta_\text{even}(t)$, respectively, decrease as $t$ increases.

\begin{figure}[t!]
\centering
\includegraphics[width=\linewidth]{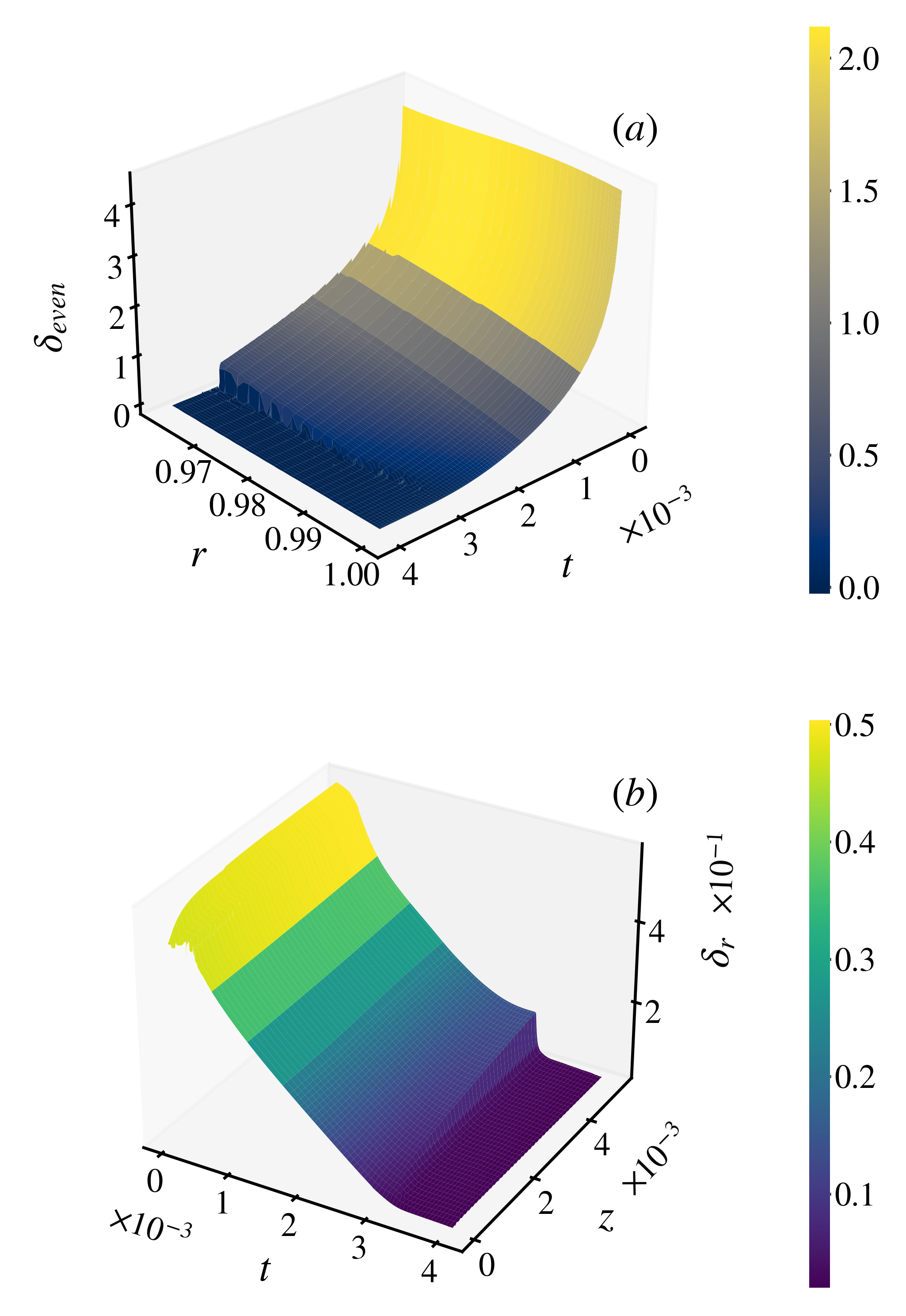}
\caption{(Color online) Surface plots of $(a)$ $\delta_{even}(r,t)$ (this falls fastest at the wall at $r=1$) and $(b)$ $\delta_{r}(z,t)$ (this falls fastest at  $z=0$).}
\label{fig:main_4}
\end{figure}

At sufficiently large $t$, $\mathcal{S}_1(r=1,k,t)$ does not have exponentially decaying
envelopes (e.g., the orange curve in Fig.~\ref{fig:main_2}), because of the
formation of tygers, our spectrally truncated system proceeds towards thermalization, 
 and we lose spectral convergence of the Fourier expansion. 

Similarly, we obtain Chebyshev spectra, at fixed values of $z$ (see panel $(b)$ in Fig. \ref{fig:main_2} and Eq. \ref{eq:S2}). At small and intermediate values of
$t$, these spectra decay exponentially at large values of $m$, with the slope 
decreasing with increasing $t$. At sufficiently large $t$,
$\mathcal{S}_2(m,z=0,t)$ does not decay at large $m$, because of the formation of tygers and the consequent loss of spectral convergence of the
Chebyshev expansion.
\section{Results}
\label{sec:Results}

\subsection{Tygers and the onset of Thermalisation}
Given the finite resolution of any practical spectral or pseudospectral DNS, we integrate not the full hydrodynamical PDE, but its Galerkin-truncated modification. 
Tygers appear when complex-space singularities come within one Galerkin
wavelength $\lambda_G = 2 \pi / k_G$ ~\cite{tyger1,tyger2,ootb,cichowlas2005effective} of the real domain. As we
increase the resolution of our DNS, $\lambda_G$ decreases, hence there is an
increase in the time taken by the pole, nearest to the real domain, to cross into
this region. Therefore, the time $t_b$ at which tygers first appear increases with
the spatial resolution of our DNS.


In the first two columns of Fig.~\ref{fig:main_3}, we present plots, versus $z$, of $u^{1}(r=1,z,t)$ [top row] and $\omega^{1}(r=1,z,t)$ [bottom row] at various times~\cite{tyger1,tyger2}; in the last column we plot  $t_{b}$, the time of the birth of tygers, versus $\lambda_G$. Tygers appear clearly in $\omega^{1}(r=1,z,t)$ before they become visible in $u^{1}(r=1,z,t)$.
{ We define tyger-birth times as the time at which oscillations, with the wavelength $\lambda_G$, are first detected by the \texttt{find\_peaks} module of MATLAB. }
 Both tyger-birth times, for the vorticity ($t_b^{(\omega^1)}$) and the velocity ($t_b^{(u^1)}$), precede (Fig.~\ref{fig:main_1}) the estimate for the singularity time given in Ref.~\cite{houluo}.
The plots in Fig.~\ref{fig:main_3} are the clearest examples of tygers in a 3D hydrodynamical PDE. 

\begin{figure}[th!]
\centering
\includegraphics[width=\linewidth]{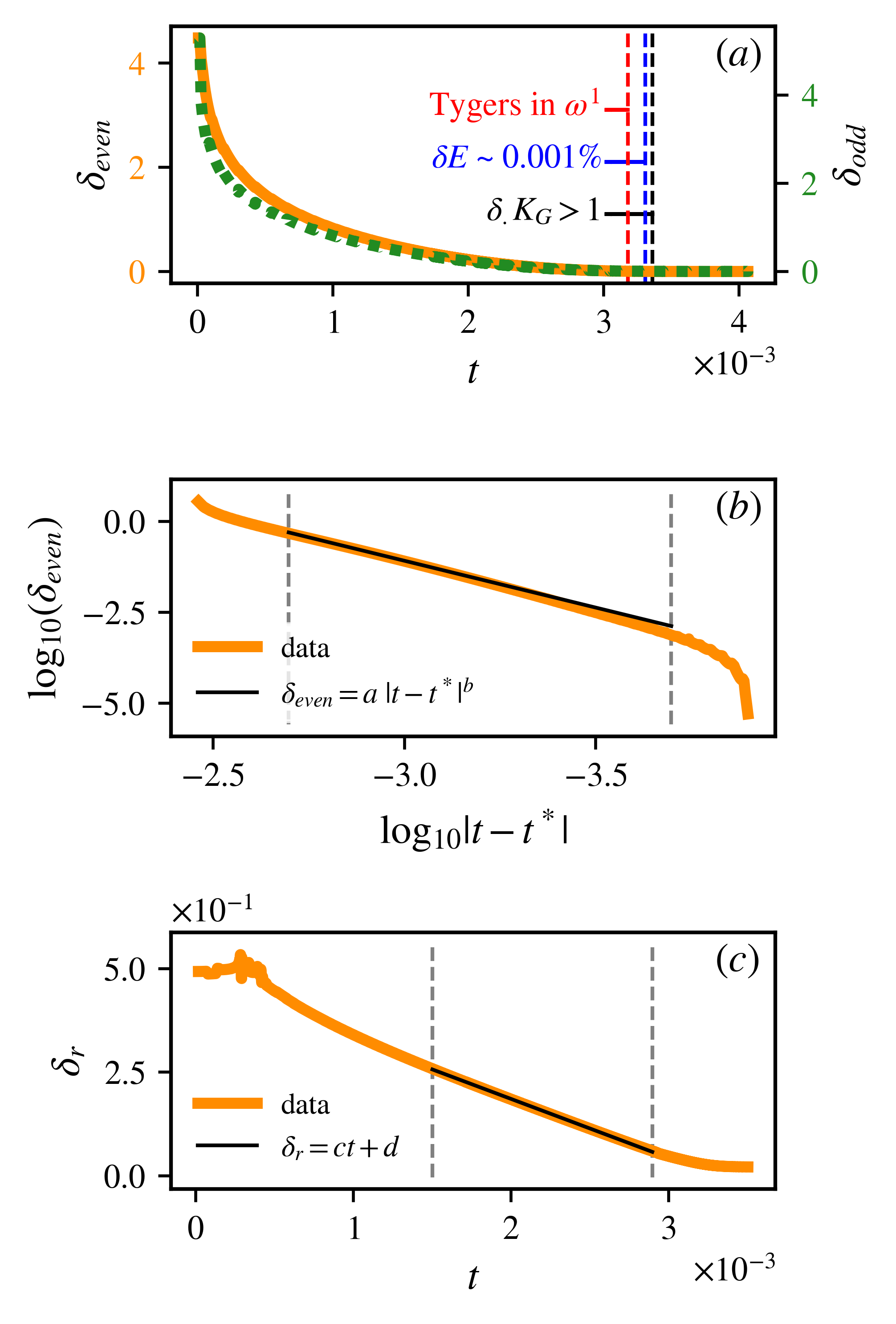}
\caption{(Color online) $(a)$ Plots versus $t$ of the widths 
$\delta_{\text{odd}}$ and $\delta_{\text{even}}$, {which we obtain from} 
the odd- and even-$k$ envelopes, respectively, of $\mathcal{S}_1(r=1,k,t)$. $(b)$ Log-log (base 10) plots of $\delta_{\text{even}}$ versus $|t-t^*|$, where $t^*=0.0035056$ is the estimate of the time of the (potential) singularity in Ref.~\cite{houluo} along with the power-law fit(black full line) $\delta_{\text{even}} = a|t-t^*|^b$. [see text]
 $(c)$ Plot of $\delta_{r}$ versus $t$ with a linear fit(black full line). [see text]
 $N_r=512$ and $N_z=4096$.}
\label{fig:main_5}
\end{figure}

As in the 1D Burgers equation~\cite{tyger1,tyger2}, tygers do not appear at the
point where the singularity develops, as a step in $u^{\theta}(r=1,z,t)$ at
$z=0$, but some distance away from it, where a resonant interaction occurs
between the fluid particle and the truncation waves~\cite{tyger1}. The tygers appear most prominently in $u_{\theta}$, which is the component of the velocity that is perpendicular to the direction in which the fastest variation in $u_{\theta}$ is seen (i.e, $\hat{z}$). The tygers
grow, as they initiate the process of thermalization and spread through the
whole domain; this is the real-space manifestation of thermalization.  The
development of the (potential) singularity leads to numerical errors as our DNS
nears the singularity-time estimate of Ref.~\cite{houluo}; eventually, energy
and helicity conservation become poor, and this prevents us from proceeding, in
our DNS, to complete thermalization.
{The plots versus $z$ in Fig.~\ref{fig:main_3} provide a natural motivation for
studying a 1D model formulated by Luo and Hou \cite{houluo}; this model displays a finite-time singularity, which we study via the analyticity-strip method and for which we show that tygers are formed before the time at which the singularity occurs (see Appendix \ref{app:1D}). }

\begin{figure*}[th]
\begin{minipage}{0.48\linewidth}
        \centering
            \includegraphics[width=\linewidth]{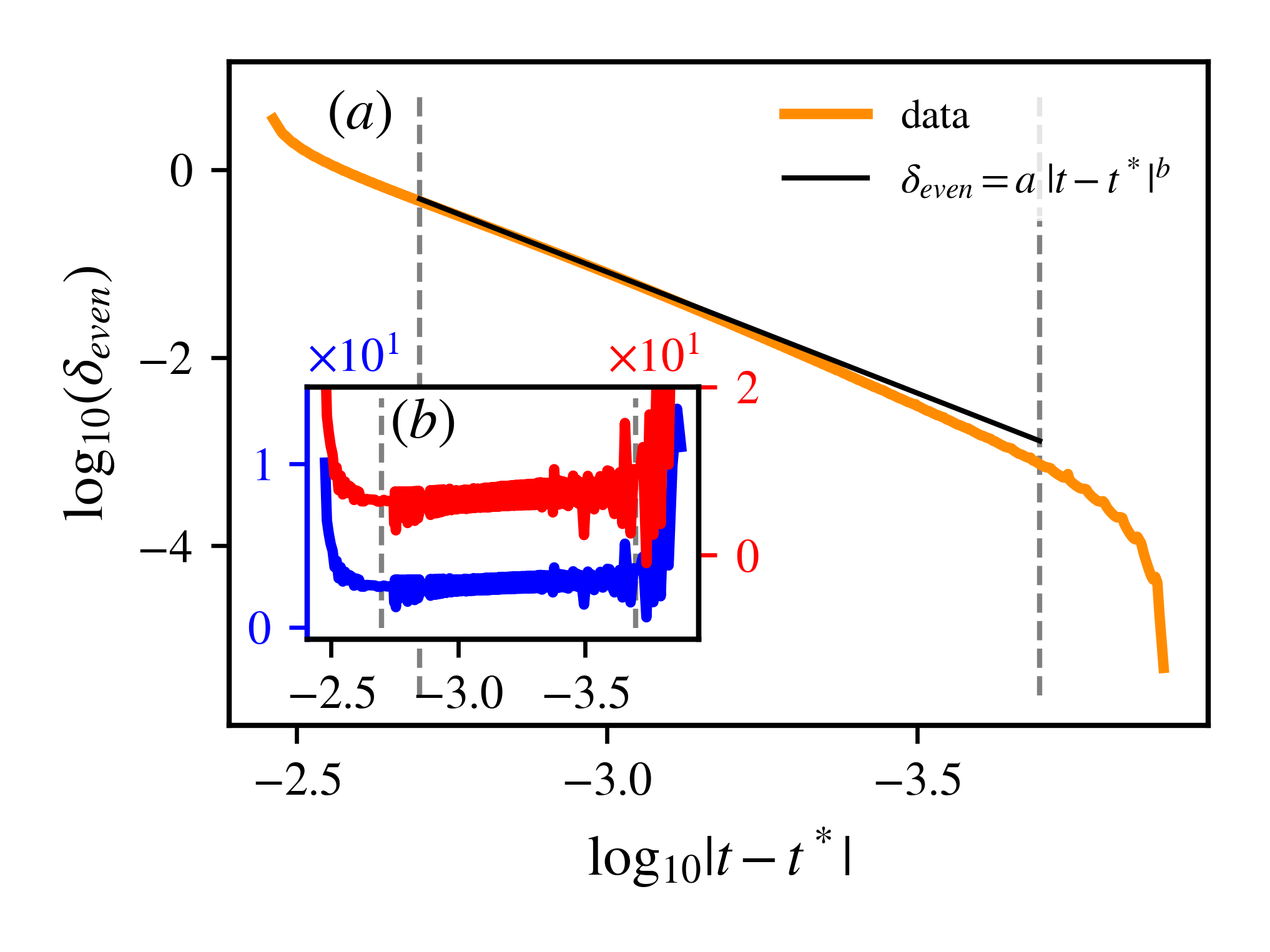}
    \caption{$(a)$ Log-log (base 10) plots of $\delta_{\text{even}}$ versus $|t-t^*|$, where $t^*=0.0035056$ along with the power-law fit(black full line) $\delta_{\text{even}} = a|t-t^*|^b$; we find $\log a= 7 \pm 1$ and $b=2.6 \pm 0.5$, in the region between the dashed grey lines. $(b)$ The fit range is based on the local-slope (blue full line) and local intercept (red full line) analysis shown in the inset panel. }
    \label{fig:supp_9a}
\end{minipage}
\hspace{1mm}
\begin{minipage}{0.48\linewidth}
    \centering
    \includegraphics[width=\linewidth]{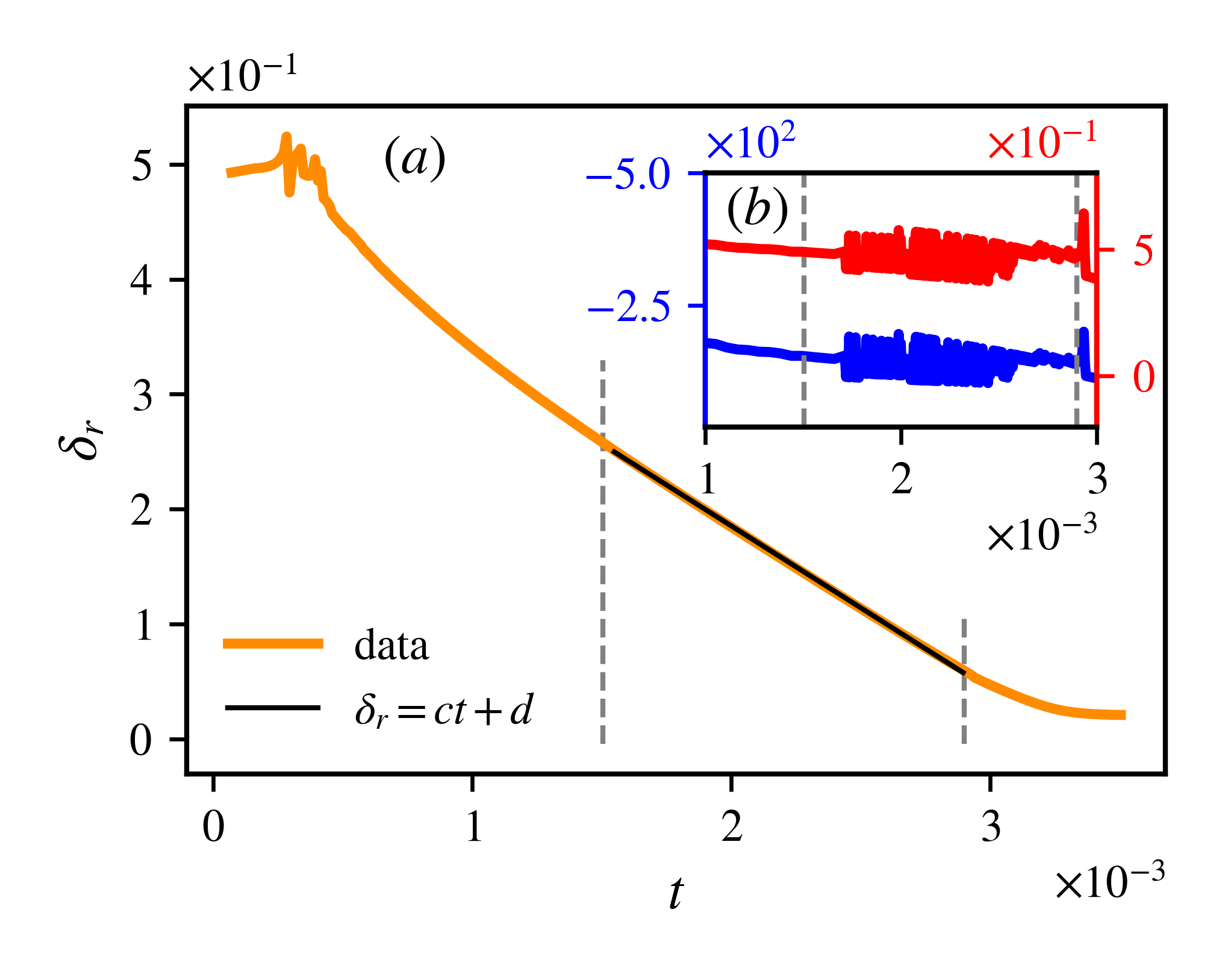}
    \caption{$(a)$ Plot of $\delta_{r}$ versus $t$; along with a linear fit (black full line) $\delta_{r} = ct+d$ with $c = -140 \pm 20  $ and $d=0.47 \pm 0.05$, in the region between the dashed grey lines. We obtain $t^* =  0.0033 \pm 0.0002$ for the x-intercept of the fit. The potential time reported by Luo et al \cite{houluo} lies in this range. $(b)$ In the inset panel, we show the local-slope (blue full line) and local intercept (red full line) analysis.}
    \label{fig:supp_9b}

\end{minipage}
\end{figure*}
\subsection{Analysis of analyticity strip widths}
The spectrum $\mathcal{S}_3(m,k,t)$, at $t=0$, has significant weight at low values of $m$ and $k$; with the passage of time, we see that this weight cascades  to large values of $m$ and $k$. This allows us to use the analyticity strip method for times when $\mathcal{S}_3(m,k,t)$ decays at large values of $m$ and $k$.

We extract $\delta_\text{even}(r,t)$ (similarly $\delta_\text{odd}(r,t)$) by using a least-squares fit for the envelopes of $\mathcal{S}_1(r,k,t)$ at even ($k_e$) and odd($k_o$) wavenumbers 
\begin{subequations}
\begin{align}
\ln(\mathcal{S}_1(r,k_{e},t)) =& C_e -n_e \ln(k_{e}) - \nonumber \\ & 2 \  \delta_\text{even}(r,t) \  k_{e}, \\
\ln(\mathcal{S}_1(r,k_{o},t)) =& C_o -n_o \ln(k_{o}) - \nonumber \\ & 2 \ \delta_\text{odd}(r,t) \  k_{o}.
\label{eq:fit_z}
\end{align}
\end{subequations}

In Fig.~\ref{fig:main_4}$(a)$, we give a surface plot of $\delta_{\text{even}}(r,t)$ to show that it decays fastest at $r=1$. Similarly, we obtain the rate at which the tail of $\mathcal{S}_2(m,z,t)$ decays exponentially, for intermediate times $t$, and thence the width $\delta_r(z,t)$ of the analyticity strip shown in Fig.~\ref{fig:main_4}$(b)$, it decays fastest at $z=0,L/2,L$. Concurrently, we see that the fastest variation in $\bom^1(r,z)$ and $\bu^1(r,z)$ occurs at the set of points corresponding to $r=1$ and $z=0,L/2,L$ where the fastest decay of the analyticity strip widths have been reported above. 

At sufficiently large $t$, there is no exponential decay (e.g., for the top plot in orange) because of the onset of thermalization in our spectrally truncated system. We use the least-squares fit
\begin{equation}
\ln(\mathcal{S}_2(m,z,t)) = C_2 - 2 m \alpha
\label{eq:fit_dr}
\end{equation}
and relate $\alpha$ to $\delta_r(z,t)$ via Eq.~(\ref{eq:alpha_dr}).


{The tygers in Fig.\ref{fig:main_3} appear as soon as the pole, in which we are
interested, enters the analyticity strip. In Fig.\ref{fig:main_4} we portray the time
dependences of the widths of analyticity strips.} We now summarise our results for analyticity-strip widths: 

In panel $(a)$ of  Fig.~\ref{fig:main_5}, we plot versus $t$, the
widths $\delta_{odd}$ and $\delta_{even}$ associated with the odd- and even-$k$ envelopes, respectively, of $\mathcal{S}_1(r=1,k,t)$. In panel $(b)$ of Fig.~\ref{fig:main_5}, we present a log-log (base 10) plot
of $\delta_{even}$ versus $|t-t^*|$, where $t^*=0.0035056$ is the
estimate of the time of the (potential) singularity in
Ref.~\cite{houluo} along with the power-law fit $\delta_{even}
= a|t-t^*|^b$;
 { we find $ \log_{10} a = 7 \pm 1 $ and $b = 2.6 \pm 0.5 $}
in the region between the dashed grey lines.
 {In Fig.~\ref{fig:supp_9a}, we show the local-slope analysis for $\log_{10}(\delta_\text{even})$ versus $\log_{10} |t-t^*|$ (of Fig. \ref{fig:main_5}(b)). We find that the slope increases linearly with time, because of the finite spatial resolution of our DNS. By using grey lines, we have indicated the region of almost constant slope, which we then use to obtain the fit in Fig. \ref{fig:main_5} $(b)$.}
{The video S3 which shows the evolution of this fit with $t$ can be found in the Supplemental Material \cite{supp}.}

 In panel $(c)$ of  Fig.~\ref{fig:main_5}, we plot, versus $t$, the width $\delta_r$, which we obtain from the natural logarithmic
decrements of $\mathcal{S}_2(m,z=0,t)$. This is very-nearly linear until just before the estimate of the time of the (potential)
singularity given in Ref.~\cite{houluo}.  From a linear fit, in
the region between the dashed grey lines, we find an
intercept, on the horizontal axis, at 
 {$t =  0.0033 \pm 0.0002$}, which is
slightly less than the estimate for the time of (potential)
singularity in Ref.~\cite{houluo}.
 {The linear fit $\delta_r = ct+b$ gives the following values for the parameters: $c = -140 \pm 20 $ and $d=0.47 \pm 0.05$.}
 {In Fig.~\ref{fig:supp_9b}, we show the local-slope analysis for $\delta_r$ versus $t$ (of Fig.\ref{fig:main_5} $(c)$). We indicate the region that is used for fitting of Fig. \ref{fig:main_5}  $(c)$ by using grey lines.}
{The video S4 which shows the evolution of this fit with $t$ can be found in the Supplemental Material \cite{supp}. }

\section{Conclusions}
\label{sec:Conclusions}
We have examined the potentially singular solution of the 3D, axisymmetric and
radially bounded Euler equation~\cite{houluo}  by developing a pseudospectral,
Fourier-Chebyshev scheme. Our method leads to new insights for it shows that,
in this scheme, the formation of tygers precedes the development of the
(potential) singularity and leads eventually to the thermalization of our
system. We then show how to generalise the analyticity-strip
method~\cite{sulem,kida1986study,ootb,bkmas,cickptg} to track this (potential)
singularity. Our results are consistent with a finite-time singularity.  A
recent paper by Barkley~\cite{barkley} has also used a Fourier-Chebyshev
method to study this initial condition; it  concentrates on the physical
mechanism for the singularity and not on the issues we discuss. Recent work by
Hertel, Besse, and Frisch~\cite{hertel}  has examined this singularity by a
Cauchy-Lagrange (CL) method, which requires the computation of Lagrangian trajectories and high-order Taylor expansions based on the Cauchy-Invariants formula; the advantage of this method is that the time step is not restricted by a Courant-Fredrichs-Lewy (CFL) criterion; however, this method is computationally expensive because it requires interpolations to map the Lagrangian grid onto the Eulerian one. This CL study also uses the BKM criterion to investigate the growth of the vorticity. Reference~\cite{houluo} uses a hybrid $6^{th}$-order Galerkin and $6^{th}$-order finite- difference method on a mesh that adapts itself in time to resolve the peak of the maximum in the vorticity (for the BKM criterion); this adaptive mesh is computationally involved and expensive. 
The smallest scale in the mesh of Ref.[18] is $\simeq 10^{- 15}$ . In our DNSs the highest resolution is $10^{-  5}$ near $r=1$, which suffices for our application of the analyticity-strip methods. Our pseudospectral method allows us to use a completely different method to track the (potential) singularity, namely, the analyticity-strip method; and given the
calculations we carry out, the CFL criterion is not a significant constraint. This singularity-detection method
gives us a complementary perspective on the development of the potential singularity that we have discussed above.

\begin{acknowledgments}
We thank SERB, CSIR, NSM, and UGC (India) and the Indo-French Centre for Applied
Mathematics (IFCAM) for their support and J.K. Alageshan, N. Besse,
M.E. Brachet, U. Frisch, A. Gupta, T. Hertel, K. Kolluru, T. Matsumoto, P.
Perlekar, S.S. Ray, and A.K. Verma for very useful discussions. We
thank, especially, N. Besse, U. Frisch, and T. Hertel for sharing the
results of their Cauchy-Lagrange study with us.   For our
high-resolution computations we have used the SahasraT CRAY computer at
the Indian Institute of Science; we thank the CRAY team here for their
support.
\end{acknowledgments}

\bibliographystyle{apsrev4-2}
\nocite{*}
\bibliography{references}

\appendix
\section{Study of the resolution dependence of the growth of $\log_{10}(\log_{10}(||\omega||_{\infty}))$}
\label{app:rescomp}

\begin{figure}[h]
\centering
\includegraphics[width=\linewidth]{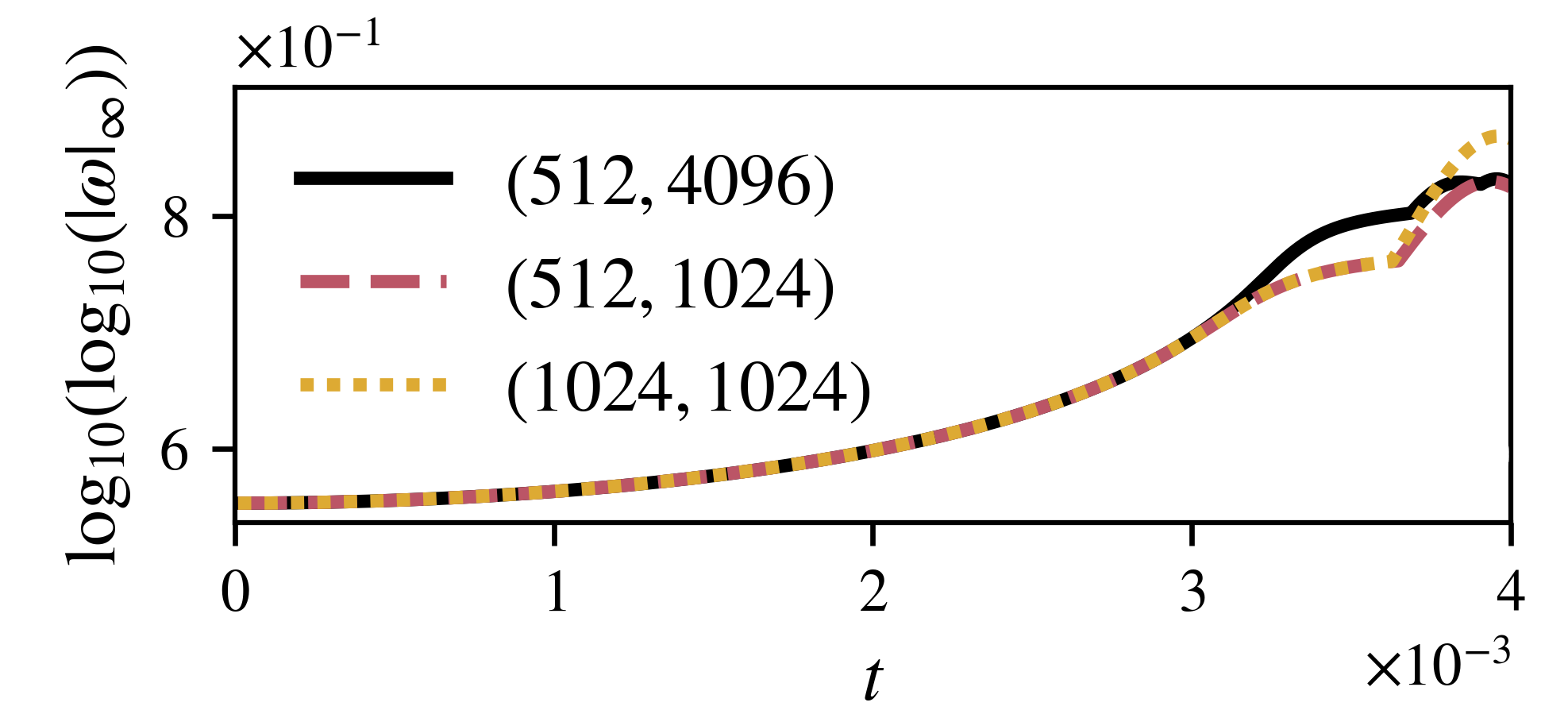}
\caption{(Color online) Plots versus time $t$ of $\log_{10}(\log_{10}(|| \omega ||_{\infty}))$, for different resolutions $(N_r,N_z)$. As $t$ increases and approaches the time of the (potential) singularity (at $t \simeq 0.0035056$), the conservation of $E$ and $H$ deteriorate. For $t \lesssim 0.0033095$, the error in energy $\delta E(t) $ is lesser than $10^{-5}\%$ for ($N_r=512,N_z=4096$). }
\label{fig:2res}
\end{figure}

In panels $(a)$ and $(b)$ of Fig.~\ref{fig:2res} we show plots of the percentage error in energy $\delta E \ \% =((E(t)-E_0)/E_0) \times 100 $ and helicity  $H$ versus time $t$ for the initial condition given by Eq.\eqref{eq:initial3D}, for the resolution $N_r=512$ and $N_z=4096$. 

The higher the resolution of our DNS (especially in the $z$ direction), the longer we can track the growth $||\omega||_{\infty}$.(as seen in panel $(c)$ of Fig.~\ref{fig:2res}). We follow the solution for as long as the percentage error in energy remains below $10^{-5}$. {Furthermore, by varying the constant factor ($100$) that multiplies the potentially singular initial condition (Eq. \eqref{eq:initial3D}), we have checked that the estimates for the blow-up and tyger-birth times are shifted to earlier times if this constant is increased. }

\begin{figure*}[ht]
\centering
\includegraphics[scale=1]{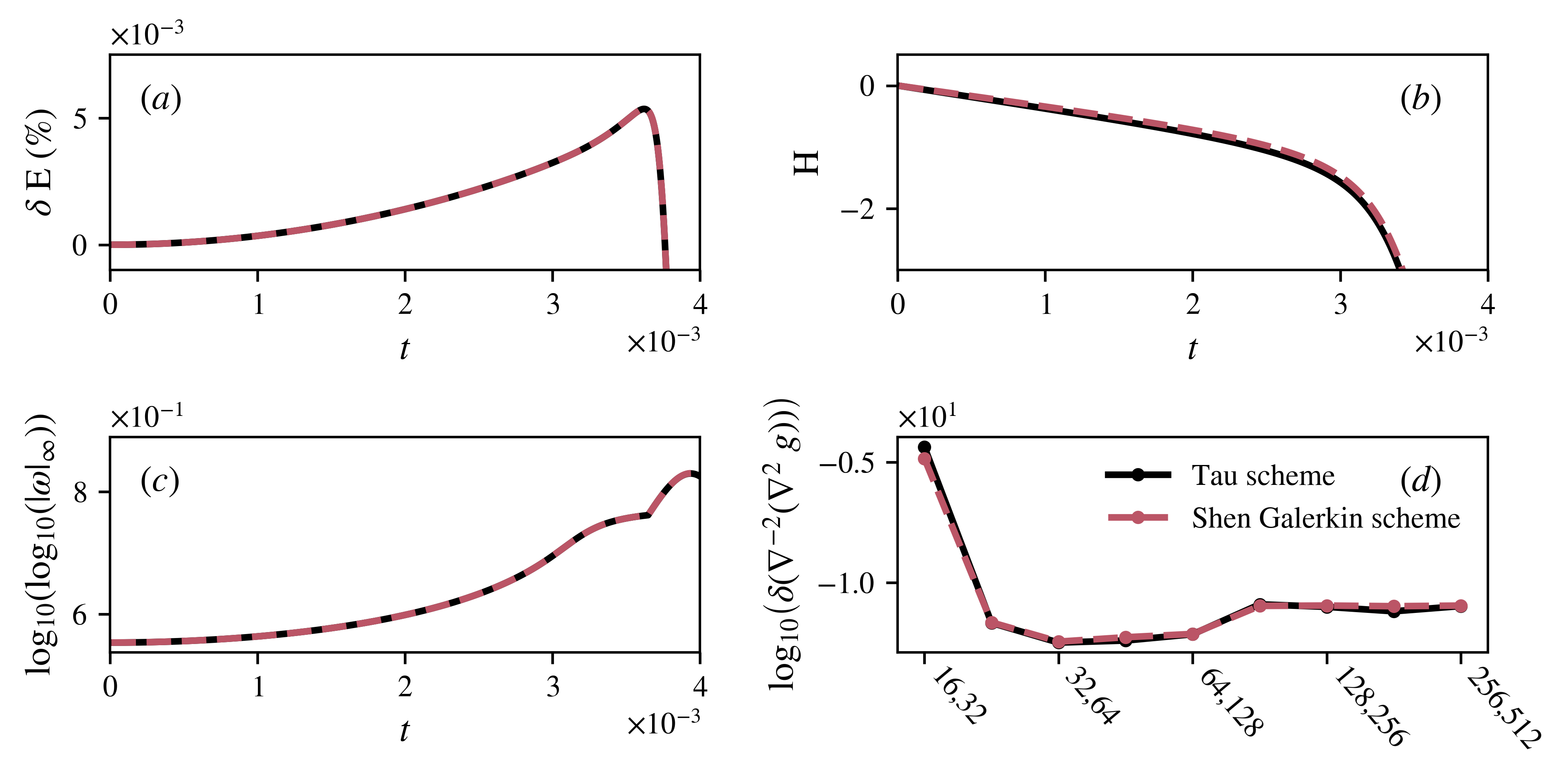}
\caption{(Color online) $(a)$Plots versus $t$ of $\log(\log(||\omega_{\infty}||))$ using the Tau and Shen-Galerkin Poisson solvers when implemented in the scheme. These plots are for a resolution of $(N_r,N_z)=(512,1024)$. $(d)$ Plots versus resolution $(N_r,N_z)$ of maximal relative error in $\nabla^{-2}(\nabla^{2} g(r,z))$. We see that both methods are equivalent (as seen by the black solid and red dashed lines that overlap almost completely).}
\label{fig:3pois} 
\end{figure*}

\section{Poisson Solvers for Axisymmetric Domains}
\label{app:poissolv}
We have checked the robustness of our results, with the Tau Poisson solver, by comparing them with those from a scheme that employs a Galerkin Poisson solver~\cite{shen1,shen2}, adapted to our boundary conditions. 

To solve Eq.\eqref{eq:main3}, we use an axisymmetric Poisson solver with the appropriate boundary conditions (Eq.\eqref{eq:noflow},\eqref{eq:polecond}) to be imposed on $\psi^1$:
\begin{subequations}
\begin{align}
 -\Big[ \ \partial_r^2 + \frac{3}{r} \ \partial_r + & \partial_z^2 \ \Big] \psi^{1}(r,z) = \omega^{1}(r,z) ;  \label{eq:poisseqn} \\
\psi^1 (1,z,t) =0; & \ \ \partial_r \psi^1 (0,z,t)=0; \\  
\psi^{1}(r,0,t) &=\psi^{1}(r,L,t).
\end{align}
\end{subequations}
Both the Shen-Galerkin and Tau methods involve the inversion of the matrix system in Eq.\eqref{eq:poisseqn} in spectral space. 

The Fourier-Chebyshev transformed system ($\partial_z^2 \rightarrow -k^2 ; r =(1+x)/2; x\in[-1,1]$) is:
\begin{subequations}
\begin{widetext}
\begin{align}
 -\Big[ \ 4  (x+1) \partial_x^2 + 12  \ \partial_x -& \ k^2 (x + 1) \ \Big] \psi^{1}(x,k) = \omega^{1}(x,k); \\
  \label{eq:poiseq_final}
 \psi^1( x=1,k)=0 \qquad &; \qquad \partial_r \psi^1(x = - 1,k)=0.
\end{align}
\end{widetext}
\end{subequations}

\subsection{Galerkin method}

This method \cite{shen1,shen2} involves the construction of basis functions $\phi_m(x)$, each of which satisfy the boundary conditions, and are  linear combinations of Chebyshev polynomials $T_m(x) = \cos(m \cos^{-1}(x) ) $:
\begin{subequations}
\begin{align}
\phi_m (x) &= T_m(x) + \frac{-4(m+1)}{(m+1)^2 + (m+2)^2} \  T_{m+1}(x) \nonumber \\  &+  \frac{m^2 + (m+1)^2}{(m+1)^2 + (m+2)^2} \ T_{m+2}(x).
\end{align}
The Galerkin approximation of $\psi^1$, in terms of $\phi_m$, is
\begin{align}
\psi^1(x,k) = \sum^{N-3}_{m=0}a(m,k)\phi_m(x).
\end{align}
We then take the weighted inner product of Eq.\eqref{eq:poisseqn} with the $\phi_m$:
\begin{eqnarray}
((x&+&1)\partial_x\psi^1,  \eta \ \phi_m ) \nonumber \\
&-& (2 \partial_x \psi^1 ,  \phi_m)_\eta + \beta ((x+1) \psi^1 , \phi_m)_\eta \nonumber \\
&=& (g,\phi_m)_\eta \,, \label{eq:Galerkinmethod}
\end{eqnarray}
\end{subequations}
where $\eta$ is the Chebyshev weight and $g = \frac{1}{4} \omega^1 (x+1)$. This matrix system can be inverted in spectral space to get $\psi^1$.

\begin{figure*}[ht]
\centering
\includegraphics[width=\linewidth]{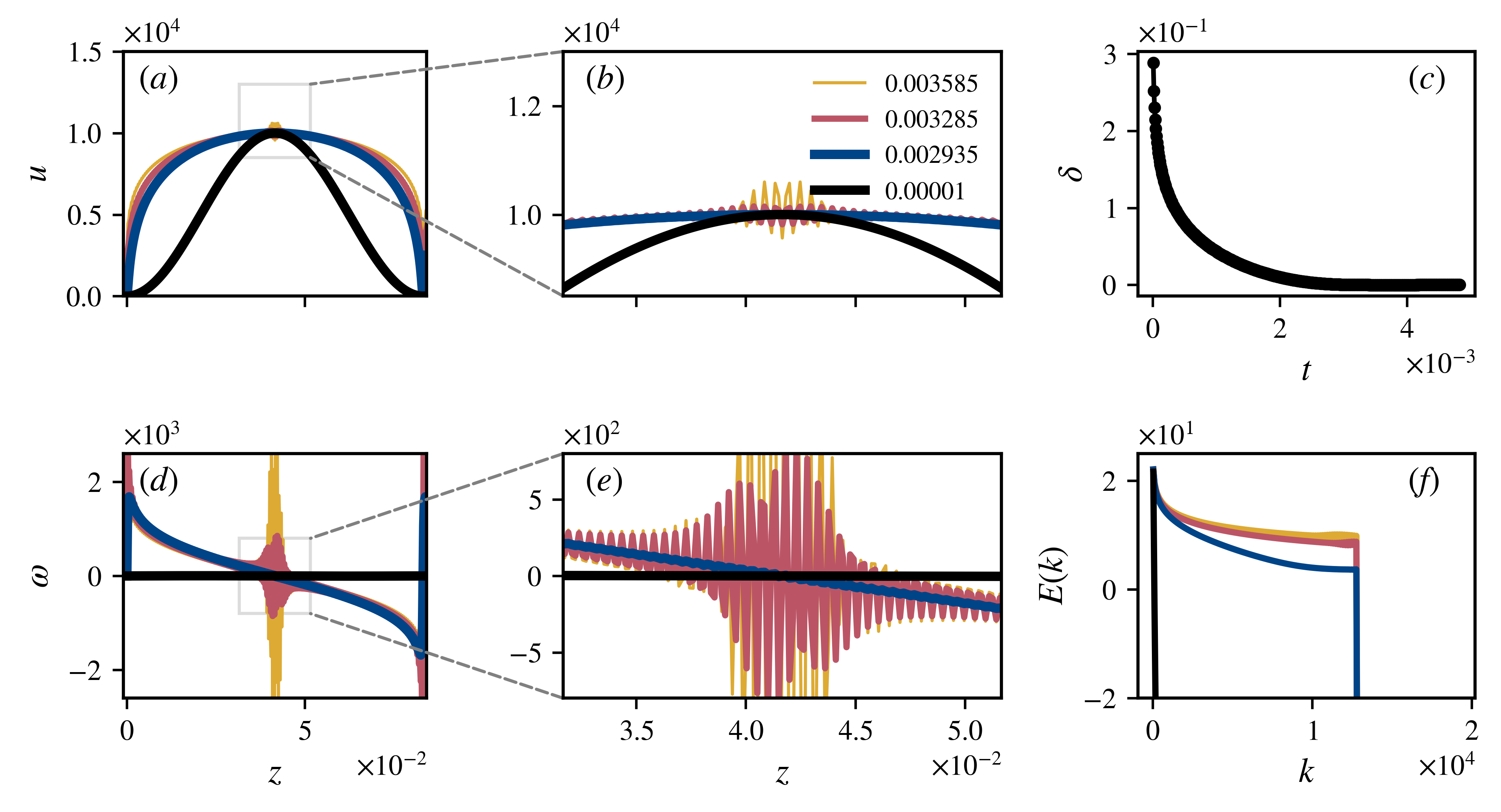}
\caption{(Color online) Plots versus $z$ of $(a),(b)$ $u$, $(d),(e)$ $\omega$; as we go from
columns one to two, we zoom in to the region with localized oscillatory structures called \textit{tygers}. $(c)$ Plots versus time $t$ of the width $\delta$ of the odd $k$ envelope of $E(k)$; $(f)$  Plots versus $k$ of $\ln(E(k))$, at different times $t$ (the full temporal evolution is given in the video S7 in the Supplemental Material \cite{supp}); $N_z=2048$; there is an exponentially decaying tail in this spectrum, at large $k$; the rate of this decay decreases with time as shown in panel $(c)$.} 
\label{fig:tw1}
\end{figure*}
\subsection{Tau method}

In this method, the boundary conditions are explicitly enforced and the basis polynomials do not satisfy the boundary conditions inherently \cite{peyret}. Here, we choose the Chebyshev polynomials as the basis;
\begin{subequations}
\begin{align}
\psi^1(x,k) = \sum^{N-1}_{m=0}a(m,k)T_m(x).
\end{align}
The weighted inner product of the Poisson equation Eq.\eqref{eq:poisseqn} is:

 \begin{eqnarray}
(-4 (x&+&1) \partial_x^2 \psi^{1},T_m)_\eta  \nonumber \\
&-& (12  \ \partial_x \psi^1, T_m)_\eta+ (k^2 (x + 1) \psi^{1},T_m)_\eta \nonumber \\
&=& (\omega^{1},T_m)_\eta . \label{eq:Galerkinmethod}
\end{eqnarray}

The last two rows of the operator matrix are replaced by the following expressions for the boundary conditions:
\begin{itemize}
\item  The no-flow boundary condition at $r=1$ :
\begin{align}
& M_{N_r-2,m} =\cos\left(2\pi m \right);  \qquad & m=0,1..N_r-1.  
\end{align}
\item The pole condition at $r=0$ is enforced as follows:
\begin{align}
 & M_{N_r-1,m} = \begin{cases}
     2m \  \sum^{m/2}_{n=1} \cos \Big( (2n-1) \pi \Big) & \\
      & m \ \text{even}; \\
     m \ \sum^{(m-1)/2}_{n=0} \cos \Big( 2n \pi  \Big) & \\
      & m \ \text{odd}.
\end{cases} 
\end{align}
\end{itemize}
\label{eq:taumethod}
\end{subequations}

Figure \ref{fig:3pois} compares the results that we obtain by using the Shen-Galerkin and Tau schemes for the Euler equation with initial condition given by Eq.\eqref{eq:initial3D}.

\section{The 1D Model}

\label{app:1D}
We have also studied the following 1D PDE, which has been introduced in Ref.~\cite{houluo} to model the potential singularity in a solution of the axisymmetric Euler equations 
restricted to $r=1$: 
\begin{subequations}
\begin{align}
\partial_t u + v \partial_z u &= 0; \\
\partial_t \omega + v \partial_z \omega &= \partial_z u;
\end{align}
\label{1deqns}
\end{subequations}
here $\partial_z v=\mathcal{H}(\omega)$, with $\mathcal{H}(.)$ the Hilbert transform; we use periodic boundary conditions~\cite{1d_1,1d_2} and the initial data 
\begin{subequations}
\begin{align}
u_0(z) &=10^4 \  \sin^2 (2 \pi z/\mathcal{L}) ; \\ \omega_0(z) &= 0.           
\end{align}
\label{1deqnsinit}
\end{subequations}

This 1D model can be obtained if we  (a) restrict the 3D axisymmetric Euler equations ~\eqref{eq:AxisymmetricEuler}
to the boundary $r=1$ and (b) then make the identifications $u(z) \rightarrow (u^1)^2(1,z)$, $\omega(z) \rightarrow \omega^1(1,z)$ and $v(z) \rightarrow \partial_r \psi^1(1,z)$. With these restrictions, the flow field is negative for $z>0$ and positive for $z<0$; this creates a compression flow at $z=0$. 
Eventually, there is a finite-time singularity in this 1D model~\cite{1d_1,1d_2}. We use a Fourier pseudospectral DNS to study this 1D model, with $N=2048$ collocation points along the $z$ axis; from this DNS we obtain the spatiotemporal evolution of $u$ and $\omega$ where $\mathcal{L} = 1/6$.

The video S5  in the Supplemental Material \cite{supp}, gives the temporal evolution of the fields and the spectra in this model. We see, once again, the development of tygers, before the time at which a finite-time singularity occurs. 
We plot these in Fig.~\ref{fig:tw1}.
The last column, top row gives a plot of the analyticity-strip width $\delta(t)$ versus the time $t$. 
The growth of tygers in this 1D model leads to thermalization in a manner that is  akin to what we have discussed for the 3D axisymmetric and radially bounded Euler (Eq.\eqref{eq:AxisymmetricEuler}); this is shown clearly by the energy spectra in the last column, bottom row of Fig.~\ref{fig:tw1}.

\section{Benchmarking of our 3D axisymmetric Euler code}
\label{app:stat}
To validate our code, we use the stationary analytical solution given in Ref.\cite{leprovost}. 
We have the following family of stationary solutions and their forms at the pole in Eqns. \eqref{eq:statsoln}.

\begin{widetext}
\begin{align}
\begin{aligned}[c]
& \psi^1 = \frac{  J_1 (\sqrt{(B^2 - \kappa^2 )} r) \ \cos(\kappa z)}{r}; \\
& u^1 = \frac{B \ J_1 (\sqrt{(B^2 -\kappa^2)} r) \ \cos(\kappa z)}{r}; \\
& \omega^1 = \frac{B^2  J_1 (\sqrt{(B^2 -\kappa^2)} r) \ \cos(\kappa z)}{r}; 
\end{aligned}
\qquad\qquad
\begin{aligned}[c]
& \psi^1(r=0) = \frac{\sqrt{B^2 - \kappa^2}}{2} \cos(\kappa z); \\
& u^1(r=0) = \frac{B \sqrt{B^2 - \kappa^2}}{2} \cos(\kappa z); \\
& \omega^1(r=0) =  \frac{B^2 \sqrt{B^2 - \kappa^2}}{2} \cos(\kappa z). 
\end{aligned}
\label{eq:statsoln}
\end{align}
\end{widetext}

 Let $x_{root}$ be one of the roots of $J_1$, then $B = \sqrt{x_{root}^2 + \kappa^2}$, where $\kappa={0,1,2..}$.

\begin{figure}[h]
\centering
\includegraphics[width=\linewidth]{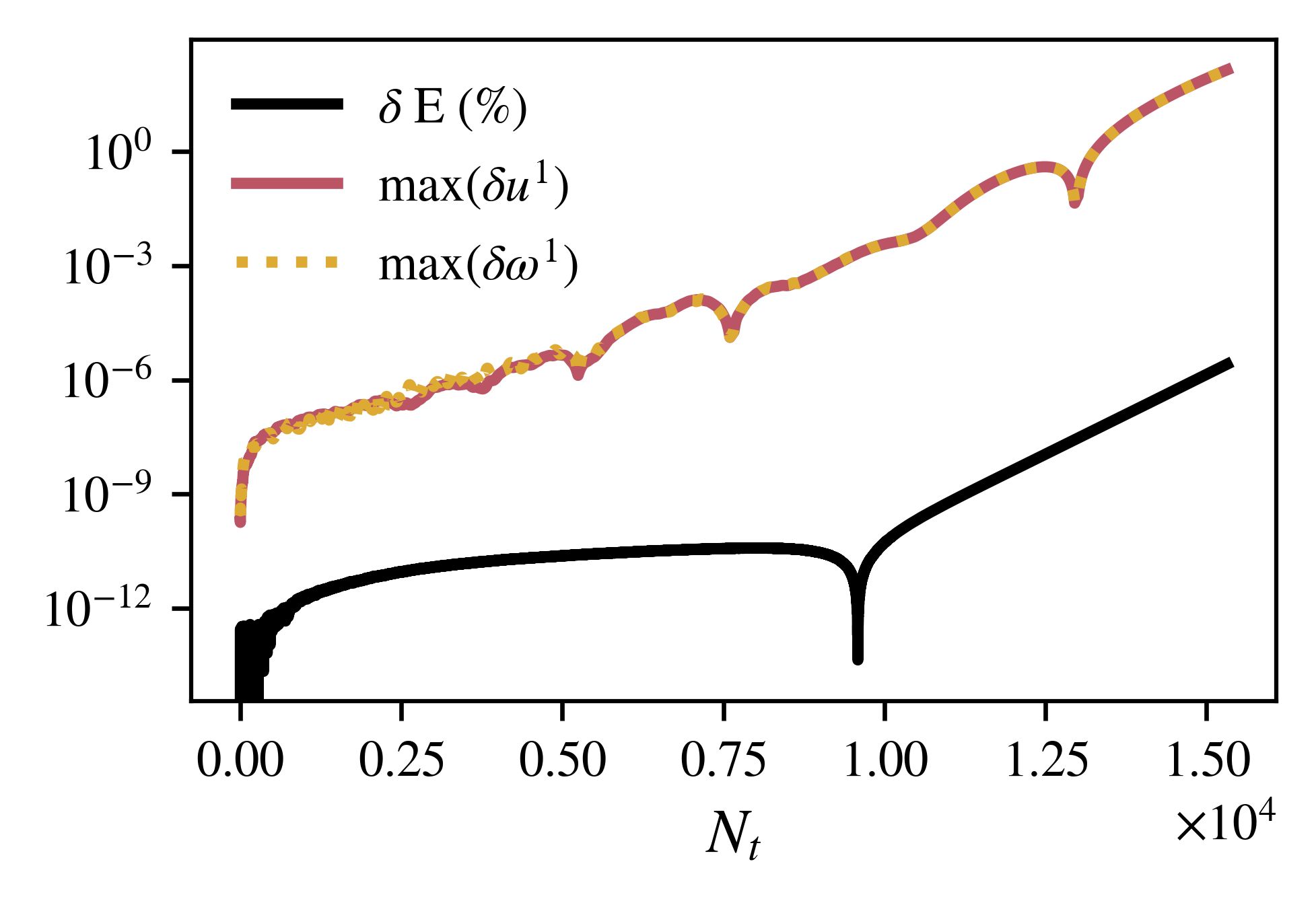}
\caption{(Color online) Plot versus $N_t$ (number of time steps) of the percentage deviation of energy $\delta E \%$, maximal relative errors in $u^1$ and $\omega^1$ for the stationary solution where $x_{root}$ is the first root of $J_1(r)$ and $\kappa=1$. }
\label{fig:SI_stat}
\end{figure}

 In Fig.\ref{fig:SI_stat}, we plot versus number of time steps $N_t$, the percentage deviation of the energy, from our DNS, relative to the energy of the stationary solution \eqref{eq:statsoln} with $\kappa=1$ and $x_{\text{root}}=3.83170597020751$ (the first root of $J_1$); the percentage deviation of energy is less than $10^{-10}$ for over $10^3$ time steps for a DNS with a resolution as low as $(N_r,N_z) = (256,512)$.

\end{document}


\preprint{APS/123-QED}

\title{Supplemental Material for: Insights from a pseudospectral study of a potentially singular solution
of the three-dimensional axisymmetric incompressible Euler equation}

\author{Sai Swetha Venkata Kolluru}
\affiliation{Centre for Condensed Matter Theory,Department of Physics, Indian Institute of Science, Bangalore, 560012, India. }
\author{Puneet Sharma}
\affiliation{Dynamics of Complex Fluids (DCF), Max Planck Institute for Dynamics and Self-Organzation, Am Fassberg 17, 37077 G\"{o}ttingen, Germany}
\author{Rahul Pandit}
\email{rahul@iisc.ac.in}
\affiliation{Centre for Condensed Matter Theory,Department of Physics, Indian Institute of Science, Bangalore, 560012, India}


\maketitle


\section{Errors due to operator subroutines}

\begin{figure}
\centering
\includegraphics[width=0.7\textwidth]{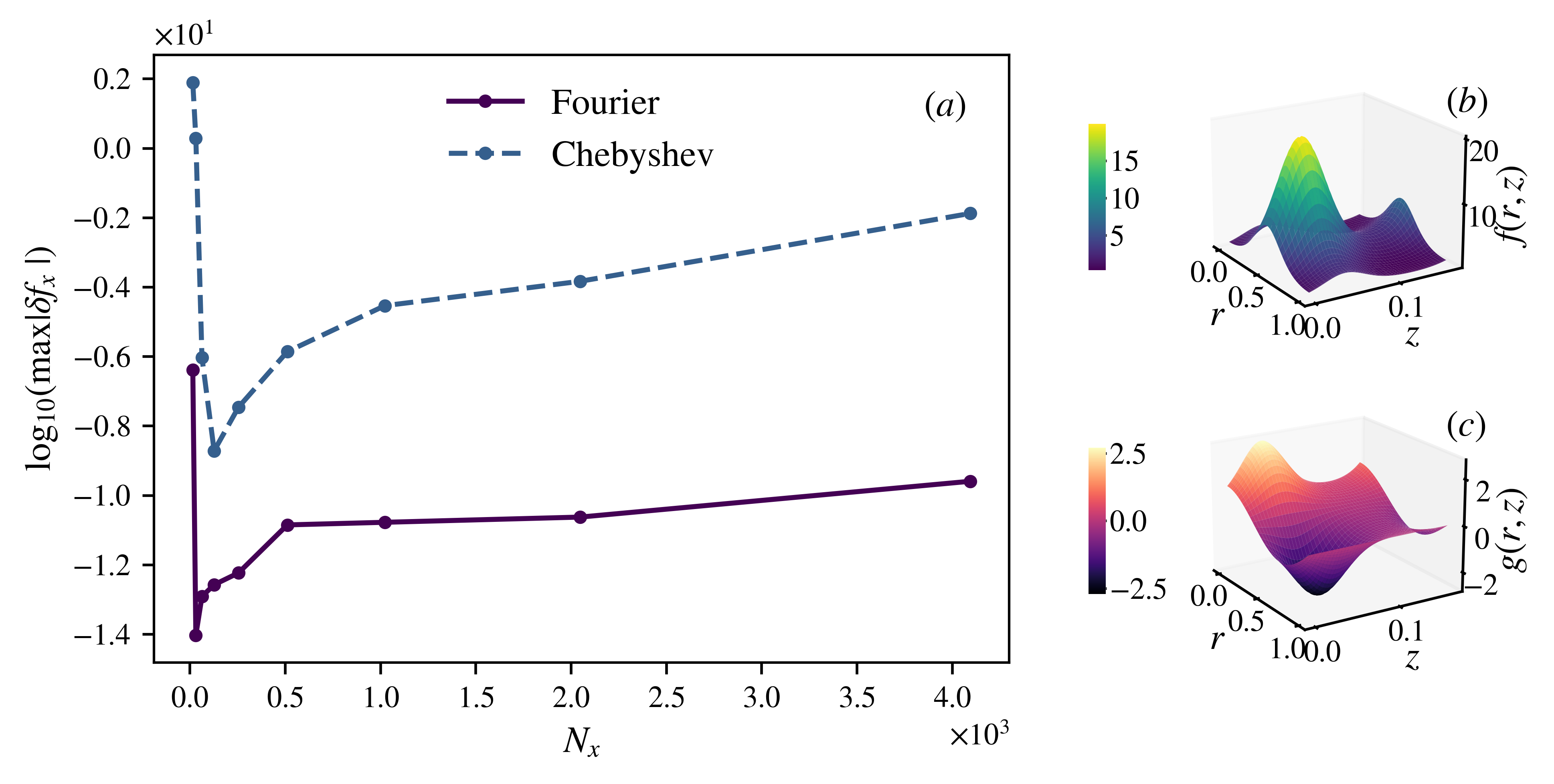}
\caption{(Color online) $(a)$ Plot versus $N_x$ (number of collocation points) of the $\log_{10} \left( \max \left| \frac{f_x }{ f_x|_\text{analytical}} -1 \right| \right)$ in the computation of the Fourier$(x=z)$ and Chebyshev$(x=r)$ derivatives for the function $f(r,z)$. Panels $(b)$ and $(c)$ show the surface plots of the test functions $f(r,z)$ and $g(r,z)$, respectively. }
\label{fig:SI_ders}
\end{figure}

We list the maximal relative errors of derivatives computed using our Fourier and Chebyshev derivative subroutines for the function: $f(r,z)= e^{1- \cos(2 \pi r)}e^{\sin(2 \pi z/L)}$(shown in panel $(b)$ of Fig.\ref{fig:SI_ders}) and  the maximal relative errors of solutions obtained from the Tau and Shen-Galerkin schemes for the Poisson problem with a source term given by  $g(r,z) = \cos(3 \pi r/2) e^{\sin(2 \pi z/L)} $(shown in panel $(c)$ of Fig. \ref{fig:SI_ders}) at $N_r =512, N_z = 4096$. We plot the maximal relative errors of Fourier and Chebyshev derivatives $f_x$ versus the number of collocation points $N_x$ in Fig.\ref{fig:SI_ders} for $x=r,z$.
\begin{itemize}
\item $\max \left| \frac{\partial_z f }{ \partial_z f|_\text{analytical}} -1 \right|= 2.5257883428768696 \times 10^{-10}$
\item $\max\left| \frac{\partial_r f }{ \partial_r f|_\text{analytical}} -1 \right|= 1.3673073130793951 \times 10^{-6}$
\item $\max\left| \frac{\nabla^{-2}(\nabla^2 g)}{ f|_\text{analytical}} -1 \right|$ for the Tau scheme=   $ 6.3672986585600666 \times 10^{-12}$
\item $\max\left| \frac{\nabla^{-2}(\nabla^2 g)}{ f|_\text{analytical}} -1 \right|$ for the Shen-Galerkin scheme =  $ 1.0352409365986301 \times 10^{-11}$
\end{itemize}

\section{Videos}
\begin{itemize}
    \item[S1] Time evolution of $\ln(\mathcal{S}_1(k,r,t))$ for $(N_r,N_z)=(256,512)$.
    \item[S2] Time evolution of $\ln(\mathcal{S}_2(m,z,t)))$ for $(N_r,N_z)=(256,512)$.
    \item[S3] Fitting $\ln(\mathcal{S}_1(k,r=1,t))$ with time $t$ for $(N_r,N_z)=(256,512)$.
    \item[S4] Fitting $\ln(\mathcal{S}_2(m,z=0,t))$ with time $t$ for $(N_r,N_z)=(256,512)$.
    \item[S5] Composite figure (in order from left to right) of the time evolution of $\omega(z)$, $\ln(E(k))$ and time series of $\delta$ of the 1-D model for $N_z =2048$.
    \item[S6] Composite figure (in order from left to right) of the time series of the helicity, time evolution of helicity isosurface in real space $H(r,z)$ and spectral space $\ln(\mathcal{S}_4(m,k,t))$, followed by the separate contributions of even and odd modes to the spectra of $\ln(\mathcal{S}_4(m,k,t))$ for $(N_r,N_z)=(256,512)$. 
    \item[S7] Composite figure (in order from left to right) of the time series of the energy, time evolution of $u^1(r,z)$, $\omega^1(r,z)$ and $\ln(\mathcal{S}_3(m,k,t))$ for $(N_r,N_z)=(256,512)$. 
\end{itemize}

\bibliography{references}